\documentstyle[epsfig]{aipproc}

\begin{document}
\title{Non-perturbative Dynamics\\ of the Heavy-Light Quark System}

\author{Nora Brambilla$^1$ and Antonio Vairo\thanks{Alexander von 
Humboldt Fellow.}}
\address{Institut f\"ur Theoretische Physik, Universit\"at Heidelberg \\
         Philosophenweg 16, D-69120 Heidelberg, FRG}

\maketitle

\begin{abstract}
We obtain the non-perturbative interaction kernel of the heavy-light 
 quark system in a gauge-invariant approach. The potential and the sum rules
 limit are discussed. Some insight is given in the light quark interaction.
\end{abstract}

\section*{Introduction}
%
%
%
%
The dynamics of a system composed by two heavy quarks is  
well understood in terms of a potential interaction (static plus relativistic 
corrections) obtained from the semirelativistic reduction of the QCD dynamics 
(see e.g. \cite{BV}).  

If at least one of the quarks is light the system behaves relativistically  
 and a pure relativistic 
treatment becomes necessary (via Dirac or Bethe--Salpeter equations).
A lot of phenomenologically justified relativistic equations 
can be found in the literature but up to now we miss a relativistic 
treatment which follows directly from QCD. Our work goes in this direction.  
In order to simplify the problem we focus on the heavy-light mesons 
in the non-recoil limit (i.e. infinitely heavy antiquark). 
Only at the end we will briefly discuss the two-body case. 
Our starting point is the quark-antiquark gauge-invariant 
Green function taken in the infinite mass limit of one particle. 
The only dynamical assumption is on the behaviour of the 
Wilson loop (i.e. on the nature of the non-perturbative vacuum). 
The gauge invariance of the formalism guarantees  
that the relevant physical information are preserved at any step 
of our derivation. In this way we obtain a QCD justified 
fully relativistic interaction kernel for the quark 
in the infinite mass limit of the antiquark. 
This kernel reduces in some region of the physical parameters 
to the heavy quark mass potential, and leads in some other region 
to the heavy quark sum rules results, providing in this 
way an unified description. We discuss our result 
with respect to the old-standing problem of the 
Lorentz structure of the Dirac kernel for a confining interaction. 
A derivation of the main results summed up here and a complete 
reference list can be found in \cite{bvplb,fsbs97}. 

\section*{The Relativistic Interaction in the One-Body limit}

The quark-antiquark Green function  
in   the Feynman--Schwinger representation 
 can be written as a quantomechanical path integral over the quark 
trajectories ($z_1(t_1)$ and $z_2(t_2)$) (quenched approximation),
\begin{eqnarray}
&~&G_{\rm inv}(x,u,y,v) =
{1\over 4} \Bigg\langle {\rm Tr}\,{\rm P}\, 
(i\,{D\!\!\!\!/}_{y}^{\,(1)}+m_1)
  \int_{0}^\infty dT_1\int_{x}^{y}{\cal D}z_1
e^{\displaystyle - i\,\int_{0}^{T_1}dt_1 {m^2+\dot z_1^2 \over 2}   }
\nonumber\\
&~& \times
\int_{0}^\infty dT_2\int_{v}^{u}{\cal D}z_2
e^{\displaystyle - i\,\int_{0}^{T_2}dt_2 {m^2+\dot z_2^2 \over 2}   }
  e^{\displaystyle ig \oint_\Gamma dz^\mu A_\mu(z)}
e^{\displaystyle i\,\int_{0}^{T_1}dt_1 {g\over 4}\sigma_{\mu\nu}^{(1)}
F^{\mu\nu}(z_1)}
\nonumber\\
&~& \times
e^{\displaystyle i\,\int_{0}^{T_2}dt_2 {g\over 4}\sigma_{\mu\nu}^{(2)}
F^{\mu\nu}(z_2)} 
(-i\,\buildrel{\leftarrow}\over{D\!\!\!\!/}_{v}^{\,(2)} + m_2) \Bigg\rangle . 
\label{Ginv2}
\end{eqnarray} 
From Eq. (\ref{Ginv2}) it emerges quite manifestly that the entire dynamics 
of the system depends on the Wilson loop
$ W(\Gamma;A) \equiv   
{\rm Tr \,} {\rm P\,} \exp \left\{ ig \oint_\Gamma dz^\mu A_\mu (z) \right\},
$
being  $\Gamma$ the closed curve defined by the quark trajectories 
and the endpoint Schwinger strings  $U(y,v)$ and $U(u,x)$.
\begin{figure}[b!] 
\vskip -1.5truecm
\centerline{\epsfig{file=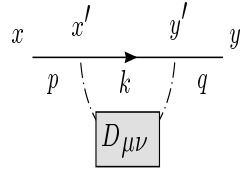,height=1.3 in,width=1.3in}}
\vspace{10pt}
\caption{The interaction kernel $K$.}
\label{figfockh}
\end{figure}
Working in  the no recoil limit  and in  the modified coordinate gauge,
 using  the cumulant expansion  \cite{DoSi} and 
keeping only bilocal cumulants we obtain:
\begin{eqnarray}
&~& \langle W(\Gamma,A) \rangle   = 
\exp \left\{ - {g^2\over 2} \int_x^y dx^{\prime\mu} \int_x^y dy^{\prime\nu}
D_{\mu\nu}(x^\prime,y^\prime) \right\},
\nonumber\\
&~& D_{\mu\nu}(x,y) \equiv
x^k y^l\int_0^1 d\alpha \, \alpha^{n(\mu)} 
 \int_0^1 d\beta\, \beta^{n(\nu)}
\langle F_{k\mu}(x^0,\alpha{\bf x})F_{l\nu}(y^0,\beta{\bf y})\rangle , 
\label{svm}
\end{eqnarray}
Assumption (\ref{svm}) corresponds to the so-called sto\-chastic vacuum 
model. Inserting expression (\ref{svm}) in Eq. (\ref{Ginv2}) 
and expanding the exponential 
taking into account only the first 
planar graph, we have
$S_D = S_0 + S_0 K S_D$  with 
 $K(y^\prime,x^\prime) = \gamma^\nu 
S_0(y^\prime,x^\prime) \gamma^\mu D_{\mu\nu}(x^\prime,y^\prime)$  
(see Fig. \ref{figfockh}. Therefore, $K$ can 
be interpreted as the interaction kernel of the Dirac equation associated 
with the motion of a quark in the field generated by an 
infinitely heavy antiquark. 
Assuming that the correlator $\langle F_{\mu\lambda}(x)F_{\nu\rho}(y)\rangle$ 
depends only on the difference between the coordinates and defining
$
 \langle F_{k\mu}(x^0,\alpha{\bf x})F_{l\nu}(y^0,\beta{\bf y})\rangle 
 \equiv f_{k\mu l\nu}(x^0-y^0, \alpha{\bf x} - \beta{\bf y}). 
$
we have 
\begin{eqnarray}
&~& K(q,p) 
 = -g^2(2\pi)\delta(p^0-q^0) 
\int_{-\infty}^{+\infty}d\tau \int_0^1 d\alpha \, \alpha^{n(\mu)}
\int_0^1 d\beta \, \beta^{n(\nu)} \nonumber\\
&\times& {\partial\over\partial p^k}{\partial\over\partial q^l}
\int d^3r \,e^{i({\bf p}-{\bf q})\cdot{\bf r}} 
\gamma^\nu 
\left\{ \theta(-\tau)\Lambda_+({\bf t})\gamma^0 e^{-i(p^0-E_t)\tau}
\right. \nonumber\\  
&-& \left.
\theta(\tau)\Lambda_-({\bf t})\gamma^0 e^{-i(p^0+E_t)\tau} \right\}\gamma^\mu
f_{k\mu l\nu}(\tau, (\alpha - \beta){\bf r}),
\label{kmom}
\end{eqnarray}
where ${\bf t} \equiv (\beta{\bf p}-\alpha{\bf q})/(\beta-\alpha)$, 
$E_t = \sqrt{t^2+m^2}$ and $\Lambda_{\pm}({\bf t}) = 
{{E_t \pm (m-{\bf t}\cdot\gamma)\gamma^0 \over 2 E_t}}$.
Focusing on the non-perturbative part, the Lorentz structure  is  
$f^{\rm n.p.}_{\mu\lambda\nu\rho}(x) = {1 \hskip -2.2 pt {\rm l}}_{\rm c} 
{\langle F^2(0) \rangle \over 24 N_c}
(g_{\mu\nu} g_{\lambda\rho} - g_{\mu\rho} g_{\lambda\nu}) D(x^2)
$
where  $\langle F^2(0) \rangle$ is the gluon condensate, 
${1 \hskip -2.2 pt {\rm l}}_{\rm c}$ the identity matrix of SU(3) and 
$D$ is a non-per\-turbative form factor normalized to unit at the origin.
Lattice simulations have shown that $D$ falls off exponentially 
(in Euclidean space-time) at long distances with a correlation 
length $a^{-1} \sim 1 ~{\rm GeV}^{-1}$.
 As shown in \cite{DoSi} this behaviour of $D$ is sufficient 
to give confinement at least in some kinematic regions. 

In what follows we study expression (\ref{kmom})   
for different choices of the parameters which are 
the correlation length $a$, the mass $m$, the binding energy 
$(p^0 - m)$ and the momentum transfer $({\bf p} - {\bf q})$. 
 
{\it A. Heavy quark potential case}  ($m>a>p^0-m$)
\noindent 
In this case with the usual reduction techniques, we obtain up to 
order $1/m^2$ the usual static and spin dependent potential \cite{BV,DoSi};
in particular for $r\to\infty$ identifying the string 
tension $\sigma = g^2 
{\langle F^2(0) \rangle \over 24 N_c} \int_{-\infty}^{+\infty} d\tau  
\int_0^\infty d \lambda \, D(\tau^2 - \lambda^2)$ 
we obtain the well-known Eichten and Feinberg result, 
$ V(r) = \sigma r -C
- {{\bf \sigma}\cdot {\bf L} \over 4 m^2} 
{\sigma \over r}, $
where $C = g^2 {\langle F^2(0) 
\rangle \over 24 N_c} \int_{-\infty}^{+\infty} d\tau  \int_0^\infty d \lambda 
\, \lambda\, D(\tau^2 - \lambda^2)$. 
We observe that the Lorentz structure  which gives origin to the negative 
sign in front of the spin-orbit potential  is in our 
case not simply a scalar ($K\simeq \sigma \, r$). 

{\it B. Sum rules case}  ($a<p^0-m$, $a<m$)
\noindent
In this case we obtain the well-known leading 
contribution to the heavy quark condensate:
$\langle \bar{Q} Q \rangle 
= -\int {d^4p \over (2\pi)^4}\int {d^4q \over (2\pi)^4} 
{\rm Tr} \left\{ S_0(q)K(q,p)S_0(p) \right\} 
= -{1\over 12} {\langle \alpha F^2(0) \rangle \over \pi m}.$

{\it C. Light quark case}  ($a>m$)
\noindent 
. Actually the case $a>m$ 
has to be considered as the only realistic one concerning 
heavy-light mesons. Under this condition either the 
exponent $(p^0 - E_t)$ as well as $(p^0 + E_t)$ can be 
neglected with respect to $a$. Therefore we have:
\begin{eqnarray}
&~& \! K(q,p) \simeq -g^2 (2\pi)\delta(p^0-q^0) 
 \int_0^{+\infty}d\tau \int_0^1 d\alpha \, \alpha^{n(\mu)}
\int_0^1 d\beta \, \beta^{n(\nu)} 
{\partial\over\partial p^k}{\partial\over\partial q^l}
\nonumber\\
&~& \! 
\times \int d^3r e^{i({\bf p}-{\bf q})\cdot{\bf r}} 
\gamma^\nu\left( 
\Lambda_+({\bf t}) - \Lambda_-({\bf t}) \right) \gamma^0 \gamma^\mu
 f^{\rm n.p.}_{k\mu l\nu}(\tau, (\alpha - \beta){\bf r}). 
\label{kmomC}
\end{eqnarray}
We observe that in the zero mass limit this expression 
gives a chirally symmetric interaction (while a purely scalar interaction 
breaks chiral symmetry at any mass scale). This means on one side that our 
interaction keeps the main feature of QCD i. e. in the zero mass limit 
chiral symmetry is broken only spontaneously. On the other side 
this seems to suggest that for very light quarks the projectors $\Lambda_+$ 
and $\Lambda_-$ which appear in (\ref{kmomC}) should be taken 
from the chiral broken solution of the corresponding Dyson--Schwinger 
equation.

\section*{Conclusion}

In the literature  a Dirac equation 
with scalar confining kernel (i. e. $K \simeq \sigma \,r$) 
has been used  to evaluate non-recoil contributions 
to the heavy-light meson spectrum. 
The main argument in favor  is the nature of the 
spin-orbit potential for heavy quarks.
 Our kernel (\ref{kmom}) follows 
simply from the assumption on the gauge fields dynamics given 
by Eq. (\ref{svm}) and by taking only one non-perturbative 
gluon insertion on the quark fermion line. 
When performing the potential reduction of this 
kernel in the heavy quark case ({\it A}) we obtain exactly 
the expected  static and spin-dependent  potentials. 
Therefore our conclusion is that there exists at least one 
non scalar  kernel which reproduces for heavy quark 
not only the Eichten and Feinberg potentials in the long distances limit, 
but also the entire stochastic vacuum model spin-dependent potential.  
Moreover when considering $a$, the inverse of the correlation length, small 
with respect to all the energy scales (case {\it B}), the kernel (\ref{kmom}) 
gives back the leading heavy quark sum rules results.  
It is possible to extend the range of applicability of 
Eq. (\ref{kmom}) to more realistic cases, like D$_{\rm s}$ and 
B$_{\rm s}$ mesons where the light quark mass is smaller than $a$
 (case {\it C}). 
The relevant part of the kernel is also in this case not a simply scalar one. 

An attempt to extend the present approach to the two-body case 
is given in \cite{fsbs97}. The equivalent graphs of Fig. \ref{figfockh}
seem to play a crucial role (in the two-body case 
such kind of graph exists for any fermion line and as exchange graph). 
Nevertheless these graphs are not sufficient in order 
to provide a complete relativistic description of the two-body system. 
The main difficulty is that in this case it does not exist 
a gauge  which automatically cancels the contributions 
coming from the end-point strings. These contributions 
are necessary in order to restore gauge invariance.

\end{document}